\documentclass[conference,final,10pt]{IEEEtran}
\usepackage[utf8]{inputenc}
\usepackage{graphicx}
\usepackage{amsmath,amssymb,amsthm,mathtools}
\usepackage{color}
\usepackage{upgreek}
\usepackage{ragged2e}
\usepackage{subfigure}
\usepackage{stfloats}
\usepackage{float}
\usepackage{bm}
\usepackage{cite}
\usepackage{cases}
\usepackage{subfigure}
\usepackage{mathbbol}
\usepackage{hyperref}
\usepackage{mathrsfs}
\usepackage{multicol}
\usepackage{algorithm, algpseudocode}

\usepackage[acronym,shortcuts]{glossaries}

\algrenewcommand\algorithmicindent{1em}
\algnewcommand\algorithmicforeach{\textbf{for each}}
\algdef{S}[FOR]{ForEach}[1]{\algorithmicforeach\ #1\ \algorithmicdo}

     \makeatletter
    \def\footnoterule{\kern-2\p@
      \noindent\hrulefill \kern 2.6\p@ \vspace{0.5ex}} 
    \makeatother

\newacronym{IoT}{IoT}{internet of things}
\newacronym{MIMO}{MIMO}{multiple-input multiple-output}
\newacronym{LRMC}{LRMC}{low-rank matrix completion}
\newacronym{MSE}{MSE}{mean square error}
\newacronym{NMSE}{NMSE}{normalized mean square error}
\newacronym{MC}{MC}{matrix completion}
\newacronym{NP}{NP}{non-deterministic polynomial-time}
\newacronym{CS}{CS}{compressed sensing}
\newacronym{SDP}{SDP}{semidefinite program}
\newacronym{RAM}{RAM}{random access memory}
\newacronym{TNN}{TNN}{truncated nuclear norm}
\newacronym{NN}{NN}{nuclear norm}
\newacronym{nmAPG}{nmAPG}{nonmonotone accelerated proximal gradient}
\newacronym{niAPG}{niAPG}{nonconvex inexact accelerated proximal gradient}
\newacronym{PG}{PG}{proximal gradient}
\newacronym{SVT}{SVT}{singular value thresholding}
\newacronym{LSP}{LSP}{log-sum-penalty}

\renewcommand{\smallskip}{\vspace{0.25cm}}

\newcommand{\norm}[1]{\left\lVert#1\right\rVert}
\newcommand{\rank}[1]{\text{rank}\left(\mathbf{#1}\right)}
\newcommand{\mask}[1]{P_\mathbb{\Omega}\left(\mathbf{#1}\right)}
\newcommand{\cmask}[1]{P_{\mathbb{\Omega}^c} \left(\mathbf{#1}\right)}

\newcommand{\tr}[1]{{\rm Tr}\left({#1}\right)}

\newcommand{\prox}[2]{\text{prox}_{#1}{\left(#2\right)}}
\newcommand{\SVT}[2]{\text{SVT}_{#1}{\left(#2\right)}}
\newcommand{\vect}[2]{\text{vec}_{#1}{\left(#2\right)}}
\newcommand{\vectinv}[2]{\text{vec}^{-1}_{#1}{\left(#2\right)}}

\begin{document}

\title{Discrete-Aware Matrix Completion via\\ Proximal Gradient}

\author{
\IEEEauthorblockN{Hiroki Iimori$^\dagger$, Giuseppe Thadeu Freitas de Abreu$^{\dagger}$, Omid Taghizadeh$^{\ddagger}$, and Koji Ishibashi$^{*}$\\}\vspace{1ex}
\IEEEauthorblockA{
$^\dagger$ Department of Comp. Sci. and Elec. Eng., Jacobs University Bremen, Campus Ring 1, 28759, Bremen, Germany \\
$^\ddagger$ Network Information Theory Group, Technische Universit{\"a}t Berlin, Straße des 17. Juni 135, 10623 Berlin, Germany \\
$^{*}$ AWCC, The University of Electro-Communications,
1-5-1 Chofugaoka, Chofu-shi, Tokyo 182-8585, Japan}
}

\maketitle

\begin{abstract}
We present a novel algorithm for the completion of low-rank matrices whose entries are limited to a finite discrete alphabet.
The proposed method is based on the recently-emerged \ac{PG} framework of optimization theory, which is applied here to solve a regularized formulation of the completion problem that includes a term enforcing the discrete-alphabet membership of the matrix entries.
\end{abstract}
\glsresetall

\section{Introduction}
\label{sect:intro}

With fair-winds of big data and \ac{IoT}, modern signal and information processing applications such as information filtering systems, networking, machine learning, and wireless communications often face a structured \ac{LRMC} problem, which intends to infer a low-rank matrix $\mathbf{X}\in\mathbb{R}^{m\times n}$ given a partially observed incomplete matrix $\mathbf{O}\in\mathbb{R}^{m\times n}$ \cite{CandesMC09,YuejieTSP19,LuongArXiv19}.
\Ac{MC} has therefore attracted much attention from both academic and industrial researchers, and has been applied to many different applications including recommender systems, localization, image compression and restoration, massive \ac{MIMO} and millimeter wave channel estimation, and phase retrieval.

To address this challenge, effective strategies based on
convex relaxation have been well-studied in the literature \cite{FazelPhD, CandesMC09, CandesSIAMOpt10} in terms of theoretical performance and complexity guarantees, of which crux is to replace the intractable non-convex rank function with its convex envelope ($i.e.,$ the \ac{NN}).
To cite several milestones, one of the earliest works \cite{FazelPhD} proposed to convert such a nuclear-norm-based optimization problem into \ac{SDP}, which however is not suitable to large-scaled problems as seen in practical scenarios due to the fact that \ac{SDP} solvers require \emph{at least} the cubic order complexity.
To circumvent this issue, the \ac{SVT} as a proximal minimizer of the \ac{NN} function was proposed in \cite{CandesSIAMOpt10}, which has been later extended to its low-complexity alternative via the Lanczos algorithm.
These methods in addition to other state-of-the-arts will be technically reviewed in Section \ref{sect:priorwork}.

In spite of intractability, structured non-convex optimization frameworks to address low-rankness have numerically shown successful performance improvements against its convex counterparts \cite{QuanmingIJCAI17}, which have recently been guaranteed to possess lower complexities from a theoretical point of view \cite{YuejieTSP19}. 
Indeed, as recently shown in \cite{LiNIPS15,QuanmingPAMI19, BalcanJMLR19}, non-convex approaches outperformed the state-of-the-art convex methods in terms of \ac{MSE} regardless of observation ratios.

Despite such intensive developments over the last decade, most of the \ac{LRMC} algorithms have been designed for general \ac{MC} problems at the cost of missing use of the most of the problem structure, leaving potential of further performance improvements.
To elaborate, many existing \ac{MC} algorithms including ones mentioned above or in Section \ref{sect:priorwork} have assumed randomness or continuity of entries of the low-rank matrix $\mathbf{X}$, albeit in many practical situations those entries must belong to a certain finite discrete alphabet set.

In this article, we therefore introduce an additive discrete-aware regularizer that can be adopted for many different state-of-the-art \ac{LRMC} algorithms, proposing a discrete-aware variate of Soft-Impute, one of the state-of-the-art methods for large-scaled \ac{LRMC} problems, so as to illustrate the effectiveness of the proposed regularizer.
Simulation results confirm the superior performance of the proposed method.

\section{Prior Work}
\label{sect:priorwork}

In this section we briefly review major \ac{LRMC} techniques studied over the last decade, which intend to recover unknown entries of a targeted low-rank matrix from partial observations, facilitating introduction to our proposed discrete-aware \ac{MC} framework.
To this end, we start with the original \ac{MC} optimization problem, which can be written as the following intractable rank minimization problem:
\begin{subequations}
\label{eqn:MC_original}
\begin{eqnarray}
{\mathop {\mathrm {argmin}} \limits_{\mathbf{X} \in \mathbb {R}^{m\times n}}} &&\!\!\!\! \rank{X}\\
\mathrm {s.t.} &&\!\!\!\! \mask{X} = \mask{O},
 \end{eqnarray}
\end{subequations} 
where $\rank{\cdot}$ denotes the rank of a given input matrix and $\mask{\cdot}$ indicates the mask operator ($i.e.,$ projection) defined as 
\begin{equation}
\left[\mask{A}\right]_{ij} = 
\begin{cases}
\left[\mathbf{A}\right]_{ij} & \text{if}\:\: (i,j)\in\mathbb{\Omega}\\
0 & \text{otherwise}
\end{cases},
\end{equation}
with $\left[\cdot\right]_{ij}$ being the $(i,j)$-th element of a given matrix and $\mathbb{\Omega}$ denoting the observed index set.

Although the global solution of equation \eqref{eqn:MC_original} corresponds to a matrix that has the lowest rank and matches observations corresponding to indexes belonging to the indicator set $\mathbb{\Omega}$, naively solving the above rank minimization problem is known to be \ac{NP}-hard due to the non-convexity of the rank operator $\rank{\cdot}$.
Taking advantage of the idea that the $\ell_0$-norm function can be replaced by its convex surrogate $\ell_1$-norm in \ac{CS}-related problems, the above rank minimization problem can be relaxed by introducing the \ac{NN} $\|\mathbf{A}\|_{*}$ ($i.e.,$ the sum of the singular values of $\mathbf{A}$) \cite{CandesMC09}, namely,
\begin{subequations}
\label{eqn:MC_nuclear}
\begin{eqnarray}
{\mathop {\mathrm {argmin}} \limits_{\mathbf{X} \in \mathbb {R}^{m\times n}}} &&\!\!\!\! \|\mathbf{X}\|_{*}\\
 \label{eqn:MC_nuclear_constraint}
\mathrm {s.t.} &&\!\!\!\! \mask{X} = \mask{O},
 \end{eqnarray}
\end{subequations} 
where note that the \ac{NN} is known to be the tightest convex lower bound of the rank operator \cite{RechtSIAM10}.

Among various numerical optimization algorithms solving equation \eqref{eqn:MC_nuclear}, one of the landmark attempts has been proposed in literature \cite{FazelPhD}, which recasts equation \eqref{eqn:MC_nuclear} as a \ac{SDP} \cite{VandenbergheSDP96}:
\begin{subequations}
\label{eqn:MC_nuclear_SDP}
\begin{eqnarray}
{\mathop {\mathrm {argmin}} \limits_{\mathbf{X}, \mathbf{W}_1, \mathbf{W}_2}} &&\!\!\!\! \tr{\mathbf{W}_1} + \tr{\mathbf{W}_2} \\
\mathrm {s.t.} &&\!\!\!\! \mask{X} = \mask{O}\\
&& \begin{bmatrix}
\mathbf{W}_1 & \mathbf{X}\\
\mathbf{X}^{\rm T} & \mathbf{W}_2
\end{bmatrix} \succeq 0
 \end{eqnarray}
\end{subequations} 
which can be solved by interior point methods available at various convex optimization solvers including SDPT3 \cite{TohOMS99}, MOSEK \cite{AndersenMOSEK00}, and SeDuMi \cite{SturmOMS99}.

Since the aforementioned \ac{SDP} solvers suffer from prohibitive time and \ac{RAM} complexity due to the nature of second-order methods, however, the above approaches are only suitable for small-sized problems in spite of the fact that we are often interested in scenarios where the dimension of $\mathbf{X}$ is large.  
Aiming at reducing the computational burden while relaxing the equality constraint \eqref{eqn:MC_nuclear_constraint} for cases where the observations contain noise or the targeted matrix to be recovered may only be regarded as approximately low-rank, various prior works including ones proposed in \cite{KeshavanTIT10, CandesSIAMOpt10, TohPJOpt10,HastieJMLR15,MazumderJMLR10, QuanmingIJCAI15, MaMP11, TaoSIAMOpt11, PrateekACM12}  can be categorized as a solution to either the problem: 
\begin{subequations}
\label{eqn:MC_nuclear_Fro_bounded}
\begin{eqnarray}
{\mathop {\mathrm {argmin}} \limits_{\mathbf{X} \in \mathbb {R}^{m\times n}}} &&\!\!\!\! \|\mathbf{X}\|_{*}\\
\mathrm {s.t.} &&\!\!\!\! \underbrace{\frac{1}{2}\|\mask{X-O}\|^2_{F}}_{\triangleq f(\mathbf{X})} \leq \varepsilon,
 \end{eqnarray}
\end{subequations} 
or its regularized form
\begin{equation}
\label{eqn:MC_nuclear_regularized}
{\mathop {\mathrm {argmin}} \limits_{\mathbf{X} \in \mathbb {R}^{m\times n}}}\:\:  f(\mathbf{X}) + \lambda \|\mathbf{X}\|_{*},
 \end{equation}
 or with the rank information
\begin{subequations}
\label{eqn:MC_nuclear_rank_info}
\begin{eqnarray}
{\mathop {\mathrm {argmin}} \limits_{\mathbf{X} \in \mathbb {R}^{m\times n}}} &&\!\!\!\! f(\mathbf{X})\\
\mathrm {s.t.} &&\!\!\!\!  \rank{X} \leq s,
 \end{eqnarray}
\end{subequations}
where $f(\cdot)$ is implicitly defined for notational convenience.

Although a great amount of efforts has been made to efficiently tackle the aforementioned convex problems\footnote{ Although equation \eqref{eqn:MC_nuclear_rank_info} is not convex, it can be efficiently solved given an accurate rank information estimate by taking advantage of the fact that the set of $s$-dimensional subspaces belonging to $\mathbb{R}^{m\times n}$ with $r\leq m$ and $r\leq n$ is a differentiable Riemannian manifold as shown in OptSpace \cite{OptSpace09Arxiv,KeshavanTIT10}}, there is growing progress on developing non-convex optimization algorithms for \ac{LRMC} as first-order methods have shown via numerical studies remarkable success in practice, which is based on non-convex regularizers such as the capped $\ell_1$-norm, the \ac{TNN}, and the \ac{LSP}.
To cite a few examples, \cite{LiNIPS15} proposed a \ac{PG} algorithm for general non-convex and non-smooth optimization, named \ac{nmAPG}, which computes gradient steps in a forward-backward fashion and is further extended in \cite{QuanmingIJCAI17} to its accelerated variate, dubbed as \ac{niAPG}.
The authors in \cite{BalcanJMLR19} study the strong duality of non-convex matrix factorization problems, proving that under certain dual conditions, the global optimality of such non-convex \ac{MC} problems can be achieved by solving its convex bi-dual problem, while \cite{SunTIT16} paves the way towards a theoretical guarantee for non-convex optimization frameworks to properly learn the targeted underlying low-rank matrix.
For more information, please refer to a recent comprehensive survey  \cite{YuejieTSP19} on non-convex \ac{MC} problems and solutions.

\section{Proposed method}
\label{sect:proposed}

As recently pointed out in \cite{LuongArXiv19}, most of the \ac{LRMC} techniques including ones mentioned above assume that entries of the targeted low-rank matrix are randomly generated ($i.e.,$ continuous random variables) in spite of the fact that many real data matrices including recommendation systems are composed of a finite set of discrete numbers, indicating potential to improve the recovery performance of the existing state-of-the-art algorithms.
To this end, in this section we introduce a discreteness-aware additive regularizer recently studied in wireless communication and signal processing literature \cite{IimoriAsilomar2019, HayakawaTWC2017,AndreiAsilomar2019,NagaharaSPL15, IimoriTWC20}, proposing a novel discrete-aware \ac{MC} algorithm as a sequence of developments \cite{QuanmingIJCAI15, QuanmingIJCAI17} stemming from Soft-Impute \cite{MazumderJMLR10}.
Notice that the proposed regularizer can be employed in various other \ac{MC} optimization frameworks, leaving such further extensions to future open problems due to the lack of space.

It is also worth noting that one may confuse the phrase ``discrete-aware'' with the existing similar research items \cite{ChenghaoAAAI19,LianSIGKDD17}, which exploit binary hashing codes for terminal user devices to reduce the storage volume and time complexity, and therefore are differentiated from the herein proposed method in the problem setup and optimization approach.
Also, the proposed approach can be differentiated from  \cite{ZhouyuanAAAI16, HuangAAAI13, NguyenSPL18} in terms of the applicability of the proposed regularizer and the optimization approach.

\subsection{Brief Summary of Soft-Impute}
Soft-Impute and its accelerated variates are state-of-the-art algorithms for large-scale \ac{LRMC} problems, which aim at solving an optimization problem similar to equation \eqref{eqn:MC_nuclear_regularized} and therefore to equation \eqref{eqn:MC_nuclear_Fro_bounded}.
To elaborate, Soft-Impute consists of the following recursion
\begin{eqnarray}\label{eqn:PG_iterates}
\mathbf{X}_{\rm t} = \SVT{\lambda}{\mathbf{X}_{\rm t-1} + \mask{O-\mathbf{X}_{\rm t-1}}},
\end{eqnarray}
where we utilized the fact that $f(\mathbf{X})$ is a convex function with $1$-Lipschitz constant, $t$ denotes the iteration index and the \ac{SVT} function is given by \cite[Theorem 2.1]{CandesSIAMOpt10} as
\begin{equation}\label{eq:SVT}
\SVT{\lambda}{\mathbf{A}} \triangleq \mathbf{U}\left(\mathbf{\Sigma} - \lambda\mathbf{I}\right)_+\mathbf{V}^{\rm T},
\end{equation}
with $\mathbf{A}\triangleq \mathbf{U}\mathbf{\Sigma}\mathbf{V}^{\rm T}$ and $(\cdot)_+$ being the positive part of the input.

It has recently been shown that Soft-Impute can be categorized as a \ac{PG} algorithm \cite{QuanmingIJCAI15}, and therefore, the well-known Nesterov-type momentum acceleration technique can be employed without loss of convergence guarantee \cite{PatrickPG2009,AntonelloArxiv2018}, leading to 
\begin{eqnarray}\label{eqn:APG_iterates}
\mathbf{X}_{\rm t} = \SVT{\lambda}{\mathbf{Y}_{\rm t} + \mask{O-\mathbf{Y}_{\rm t}}},
\end{eqnarray}
with $\mathbf{Y}_{\rm t} \triangleq (1+ \beta_{\rm t})\mathbf{X}_{\rm t-1} + \beta_{\rm t} \mathbf{X}_{\rm t-2}$ where $\beta_{\rm t}$ is the momentum weight.

\subsection{Discrete-Aware Matrix Completion}
Assuming that entries of the matrix to be recovered belong to a certain finite discrete alphabet set $\mathbb{A}\triangleq \{a_1,a_2,\cdots\}$ ($e.g.,$ integers in case of recommendation systems), we intend to tackle a variety of the following regularized minimization problem
\begin{equation}
\label{eqn:MC_proposed}
{\mathop {\mathrm {argmin}} \limits_{\mathbf{X} \in \mathbb {R}^{m\times n}}}\:\:  f(\mathbf{X}) + \lambda g(\mathbf{X}) + \xi r(\mathbf{X}|p),
 \end{equation}
where $g(\mathbf{X})$ denotes a non-smooth (possibly non-convex) low-rank regularizer \cite{QuanmingPAMI19}, $\xi \geq 0$, and 
\vspace{-1ex}
\begin{equation}\label{eq:DAreg}
r(\mathbf{X}|p) \triangleq \sum^{|\mathbb{A}|}_{k=1}\|\vect{\mathbb{\Omega}^c}{\mathbf{X}} - a_k\mathbf{1}\|_p\vspace{-1ex}
\end{equation}
where $r(\mathbf{X}|p)$ is the discrete-space regularizer\footnote{\color{black}Although it has been shown in the literature \cite{IimoriAsilomar2019, HayakawaTWC2017, IimoriICOIN20, NagaharaSPL15, IimoriTWC20,HayakawaAccess2018,AndreiAsilomar2019} that the base of the norm function is set to be $p=0$ or $p=1$ to enhance the discreteness of the inputs, the base $p$ can be any positive number in principle.} with $0\leq p$, $\vect{\mathbb{\Omega}^c}{\mathbf{X}}$ denoting vectorization of entries of $\mathbf{X}$ corresponding to a given index set $\mathbb{\Omega}^c$, and $\mathbb{\Omega}^c$ being the complementary set of $\mathbb{\Omega}$.

Although non-convex scenarios where either $g(\mathbf{X})$, $r(\mathbf{X}|p)$ or both are non-convex regularizer(s) can be considered, we hereafter focus on the convex scenario ($i.e.,$ $g(\mathbf{X})= \|\mathbf{X}\|_{*}$ and $r(\mathbf{X}|1)= \sum^{|\mathbb{A}|}_{k=1}\|\vect{\mathbb{\Omega}^c}{\mathbf{X}} - a_k\mathbf{1}\|_1$) for the sake of simplicity and because of space constraints\footnote{A full description of algorithmic designs with non-convex regularizers will be given together with the associated pseudo-codes in the final version of the manuscript.}.
The accelerated \ac{PG} algorithm for a discrete-aware convex variate of Soft-Impute as described in equation \eqref{eqn:MC_proposed}, which hold the convergence rate $\mathcal{O}\big(\frac{1}{t^2}\big)$, can be summarized as the following recursion:
\begin{subequations}\label{eq:proposed_convexAPG}
\begin{eqnarray}
\mathbf{Y}_{\rm t} &=& (1+ \beta_{\rm t})\mathbf{X}_{\rm t-1} + \beta_{\rm t} \mathbf{X}_{\rm t-2}\\
\mathbf{Z}_{\rm t} &=&  \prox{\xi r}{\mathbf{Y}_{\rm t}} \\
\mathbf{X}_{\rm t} &=& \SVT{\lambda}{\cmask{\mathbf{Z}_{\rm t}}+\mask{O}}
\end{eqnarray}
\end{subequations}
where $ \prox{\xi r}{\mathbf{Y}_{\rm t}}$ is the proximal operator given by
\begin{equation}\label{eq:prox}
 \prox{\xi r}{\mathbf{Y}_{\rm t}} \triangleq \underset{{\mathbf{U}}} {\mathrm{argmin}}\:\:\:  r(\mathbf{U}|1) + \frac{1}{2\xi}\norm{\vect{\mathbb{\Omega}^c}{\mathbf{U}-\mathbf{Y}_{\rm t}}}^2_{2}.\!\!
\end{equation}

Taking into account the fact that the proximal operator of a sum of convex regularizers can be computed from a sequence of individual proximal operators \cite{PatrickPG2009}, we readily obtain
\begin{equation}\label{eqn:prox_sum}
 \prox{\xi r}{\mathbf{Y}_{\rm t}} = \prox{\xi r_1}{\prox{\xi r_2}{\cdots\prox{\xi r_{|\mathbb{A}|}}{\mathbf{Y}_{\rm t}}}},
\end{equation}
where $r_k(\mathbf{Y}_{\rm t})\triangleq \|\vect{\mathbb{\Omega}^c}{\mathbf{Y}_{\rm t}} - a_k\mathbf{1}\|_1$ for $k\in\{1,2,\ldots, |\mathbb{A}|\}$.

To this end, each proximal operator can be written as
\begin{equation}\label{eq:prox_each}
 \prox{\xi r_k}{\mathbf{Y}_{\rm t}} \triangleq \underset{{\mathbf{U}}} {\mathrm{argmin}}\:\:  \|\mathbf{u} - a_k\mathbf{1}\|_1 + \frac{1}{2\xi}\norm{\mathbf{u}-\mathbf{y}_{\rm t}}^2_{2},
\end{equation}
with $\mathbf{u}\triangleq \vect{\mathbb{\Omega}^c}{\mathbf{U}}$ and $\mathbf{y}_t\triangleq \vect{\mathbb{\Omega}^c}{\mathbf{Y}_t}$, which can be compactly written element-by-element as
\begin{equation}\label{eq:prox_each_element}
\underset{\bar{u}_\ell} {\mathrm{argmin}}\:\:  |\bar{u}_\ell| + \frac{1}{2\xi}(\bar{u}_\ell- \bar{y}_{{\rm t},\ell})^2,
\end{equation}
where $\bar{u}_\ell\triangleq \left[\mathbf{u}\right]_\ell - a_k$, $\bar{y}_{{\rm t},\ell}\triangleq \left[\mathbf{y}_t\right]_\ell - a_k$, $\bar{\mathbf{u}}\triangleq [\bar{u}_1, \bar{u}_2, \ldots, \bar{u}_{ |\mathbb{\Omega}^c|}]^{\rm T}$, $\bar{\mathbf{y}}_{\rm t} \triangleq [\bar{y}_{{\rm t},1}, \bar{y}_{{\rm t},2}, \ldots, \bar{y}_{{\rm t}, |\mathbb{\Omega}^c|}]^{\rm T}$ and $\ell\in\{\mathbb{Z}| 1\leq\ell\leq |\mathbb{\Omega}^c|\}$.

One readily notice that equation \eqref{eq:prox_each_element} has a closed form solution ($i.e.,$ soft-thresholding function) given by
\begin{equation}\label{eqn:softthr}
\bar{\mathbf{u}} = \text{sign}\left(\bar{\mathbf{y}}_{\rm t} \right)\odot(|\bar{\mathbf{y}}_{\rm t}| - \xi\mathbf{1})_+
\end{equation}
where $\odot$ is the Hadamard product and $\text{sign}\left(\cdot \right)$ denotes the (element-wise) sign function.

Notice that in equation \eqref{eqn:softthr}, $|\bar{\mathbf{y}}_{\rm t}|$ performs the element-wise absolute operation.
Finally we recover $\mathbf{u}$ by
\begin{equation}
\mathbf{u} = \bar{\mathbf{u}} + a_k\mathbf{1},
\end{equation}
and $\mathbf{U}$ by mapping $\mathbf{u}$ onto the unobserved indexes, namely,
\begin{equation}\label{eq:InvVec}
\mathbf{U} = \vectinv{\mathbb{\Omega}^c}{\mathbf{u}},
\end{equation}
where $\vectinv{\mathbb{\Omega}^c}{\cdot}$ denotes the inverse function of $\vect{\mathbb{\Omega}^c}{\cdot}$.

\section{Numerical Evaluation}
\label{sect:results}

\begin{figure}[t!]
\includegraphics[width=\columnwidth]{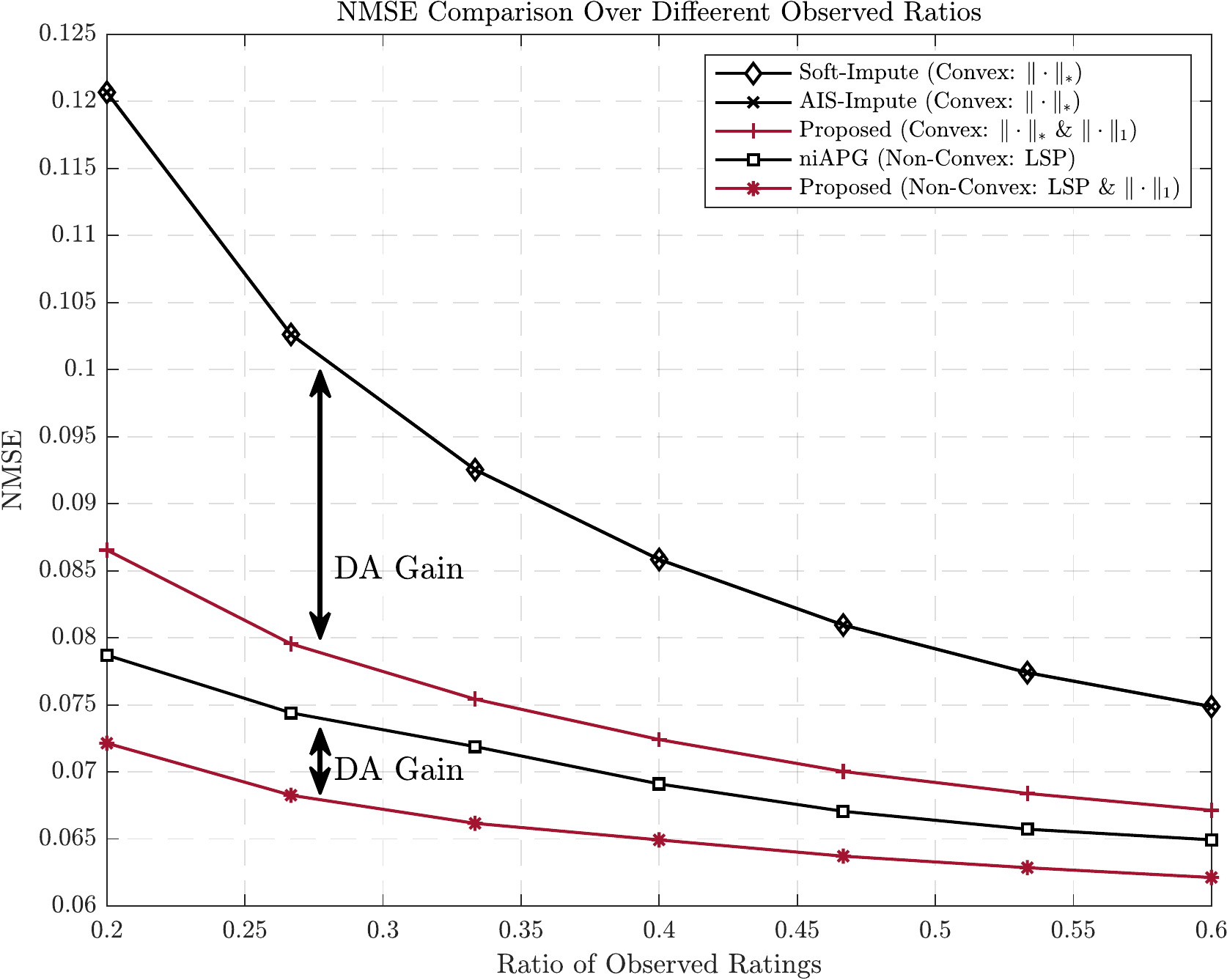}
\caption{\ac{NMSE} performance evaluations of the proposed discrete-aware \ac{MC} algorithms (red) and other first-order state-of-the-art methods (gray and black) with respect to different observation ratios.}
\label{fig:NMSE}
\vspace{1ex}
\includegraphics[width=\columnwidth]{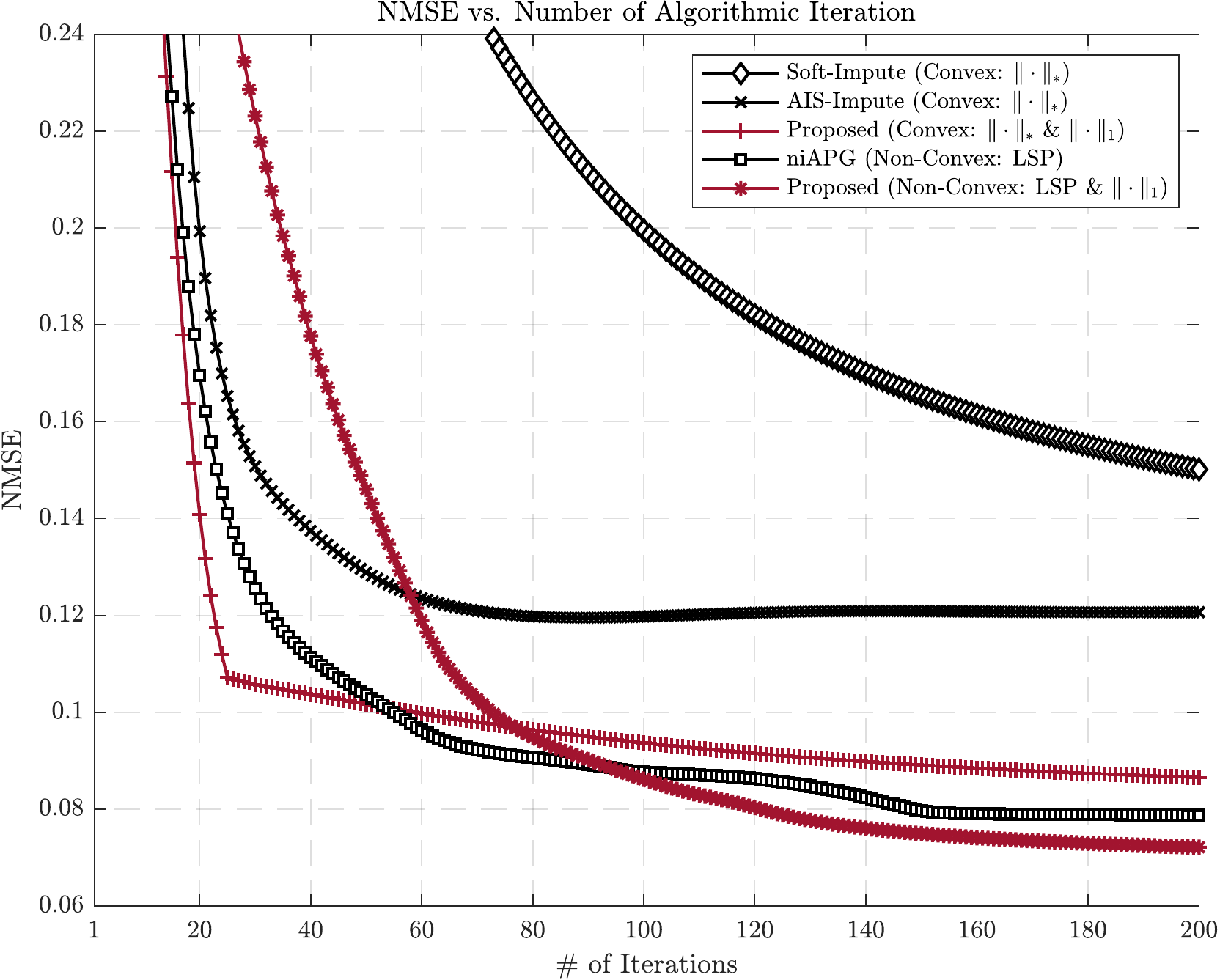}
\caption{\ac{NMSE} performance behavior on the MovieLens-100k data set as a function of algorithmic iterations at $20$\% observed ratio.}
\label{fig:Convergence}
\end{figure}

In this section, we perform numerical experiments on discrete-valued real-world data sets to evaluate the proposed discrete-aware \ac{MC} algorithm.
\textcolor{black}{To this end, we adopt the MovieLens-100k data set\footnote{https://grouplens.org/datasets/movielens/} for recommender systems, one of the popular data sets utilized in \ac{MC} literature for performance evaluations, which is composed of integer ratings (from $1$ to $5$) associated with many different user-movie pairs and possesses a low-rank nature due to the inter-user correlation in preferred movies.}
To evaluate the robustness jointly with the recovery performance, we vary the observed ratio from $20$\% to $60$\%, while \ac{NMSE} is utilized as the performance metric, which is given by 
\begin{equation}
\text{\ac{NMSE}} \triangleq \frac{\|\cmask{X-O}\|^2_F }{\|\cmask{O}\|^2_F}.
\end{equation}

Besides the Soft-Impute algorithm \cite{MazumderJMLR10}, we compare our proposed algorithm with other state-of-the-art methods such as AIS-Impute \cite{QuanmingIJCAI15}, an accelerated variate of Soft-Impute, niAPG \cite{QuanmingIJCAI17}, a non-convex variate of Soft-Impute with the \ac{LSP} non-convex regularizer.

The \ac{NMSE} performance results comparing our proposed discrete-aware variates of Soft-Impute and the aforementioned state-of-the-art \ac{LRMC} algorithms as a function of ratio of observed ratings for training are shown in Figure \ref{fig:NMSE}, where the red lines correspond to our proposed methods and black or gray lines are associated with the state-of-the-arts.
For the sake of clarity, the \ac{NMSE} performance gaps due to discreteness-awareness is highlighted by annotation arrows.
It can be observed from the figure that most of the algorithms are able to successfully achieve less than $0.1$ in terms of \ac{NMSE} for a wide range of observed ratios, albeit non-convex algorithms with \ac{LSP} can reduce the performance degradation in a severe scenario where only a few number of entries of the matrix can be observed.
More interestingly, even in case of convex algorithms, the discreteness-awareness considerably decrease increment of the \ac{NMSE} curve at the low observed ratio range, which indicates the robustness of the proposed discrete-aware regularizer.

In Figure \ref{fig:Convergence}, the \ac{NMSE} convergence behavior of the algorithms with respect to the number of algorithmic iterations is presented, where we can perceive that most of the algorithms converge within $100$ iterations in case of with the convex \ac{NN} regularizer and $180$ iterations in case of with the non-convex \ac{LSP} regularizer, respectively. 
Furthermore, the figure illustrates the accelerated convergence of the proposed algorithm with the convex \ac{NN} regularizer.
According to this observation, it may be concluded that the discreteness-awareness can not only improve the \ac{NMSE} performance but also contribute to finding the optimality condition.
However, the latter benefit is not necessary in case of non-convex scenarios due to multiple local minima, which rather results in slightly slower convergence.

{\color{black} Besides the above, we remark that the additional complexity due to the discreteness-aware regularizer in equation \eqref{eq:DAreg} with $p=1$ is linear with respect to the cardinality of the unknown index set ($i.e.,$ $|\mathbb{\Omega}^c|$) as one may readily observe from the element-by-element operation in equations \eqref{eqn:softthr}--\eqref{eq:InvVec}.
Therefore, one may conclude that the most expensive part of the algorithm in terms of complexity is the same as that of the state-of-the-art methods, $i.e.,$ \ac{SVT}, indicating that the proposed algorithm maintains the same complexity order.

In case of $p=0$, however, the regularizer may affect the convergence or the complexity of the proposed \ac{PG} algorithm due to many different reasons such as expansiveness of $r_0(\mathbf{X})$ \cite{BurgerSMCISE17} or successive convex approximation to relax the $\ell_0$-norm function.
Taking into account the aforementioned issues, we will provide a \ac{PG}-based algorithm for $r_0(\mathbf{X})$ in the final version of the manuscript due to the space limitation at the submission.

In light of all the above, we conclude from the numerical performance evaluations that our proposed discreteness-aware \ac{MC} algorithm may further accelerate the convergence and improve the completion performance in case of adopting convex functions for both regularizers ($i.e.,$ $g(\cdot)=\|\cdot\|_*$ and $r_1(\cdot)$), while enjoying the uniqueness of the solution due to the convexity of equation \eqref{eqn:MC_proposed}. 
In case of non-convex low-rank regularizer ($i.e.,$ \ac{LSP}) while maintaining convex discreteness-aware regularizer ($i.e.,$ $r_1(\cdot)$), it has been shown that at the expense of slower convergence, the \ac{NMSE} performance can be enhanced as shown in Figure \ref{fig:Convergence}.
}

\section{Conclusion and Remarks}
\label{sect:conclusion}

In this article, we proposed a novel discrete-aware \ac{LRMC} algorithm for structured practical \ac{MC} problems where entries of the matrix to be recovered is subject to a certain finite discrete alphabet set such as recommender systems.
To tackle this open problem indicated by a recent comprehensive survey \cite{LuongArXiv19}, we introduce a discrete-aware additive regularizer that has been recently considered in signal processing and compressive sensing literature.
Performance evaluations via software simulations demonstrate the superior performance of the proposed methods due to the awareness to such specific structure in the targeted matrix.
We conclude this article by providing some possible applications of the proposed \ac{MC} algorithm. 

\begin{itemize}
\item The most important and obvious application is recommender systems for Netflix, Amazon, and so on with discrete scores, e.g., 1 of 5. 
\item Another application of the proposed MC algorithm is an estimation of connections among users in networks such as social networks, large wireless ad-hoc networks, etc where 0 means not-connected and 1 denotes connected. Although it is not impossible to obtain the whole adjacency matrix from the network, it would cost tons of resources for that. For example, if the proposed MC precisely estimates the whole matrix of the wireless ad-hoc network with the partial information, it enables the network to perform the optimal routing, network coding, distributed coding, and so on with even less overhead, which results in significant improvement of the throughput.
\item  An interesting field to which discreteness-aware \ac{MC} algorithms can be applied is an index coding problem in a broadcast channel \cite{EsfahanizadehIITW14}, where a single source communicates with multiple-users over a rate-limited channel.
\end{itemize}

\vspace{-1ex}
\bibliographystyle{IEEEtran}
\bibliography{listofpublications}

\end{document}